\documentclass[zpreprint,zbstdefault]{zeus_paper}
\usepackage[english]{babel}
\chardef\usc=95
\chardef\til=126
\catcode`\@=11 
\DeclareRobustCommand\xdotspace{\futurelet\@let@token\@xdotspace}
\def\@xdotspace{%
  \ifx\@let@token.\else
  \ifx\@let@token\bgroup.\else
  \ifx\@let@token\egroup.\else
  \ifx\@let@token\/.\else
  \ifx\@let@token\ .\else
  \ifx\@let@token~.\else
  \ifx\@let@token!.\else
  \ifx\@let@token,.\else
  \ifx\@let@token:.\else
  \ifx\@let@token;.\else
  \ifx\@let@token?.\else
  \ifx\@let@token/.\else
  \ifx\@let@token'.\else
  \ifx\@let@token).\else
  \ifx\@let@token-.\else
  \ifx\@let@token\@xobeysp.\else
  \ifx\@let@token\space.\else
  \ifx\@let@token\@sptoken.\else
   .\space
   \fi\fi\fi\fi\fi\fi\fi\fi\fi\fi\fi\fi\fi\fi\fi\fi\fi\fi}
\catcode`\@=12 

\newcommand{\stru}[2]{%
   \relax\ifmmode\hbox{\vrule height#1 depth#2 width0pt}%
   \else\vrule height#1 depth#2 width0pt\fi}

\newcommand{\Ronum}[1]{\uppercase\expandafter{\romannumeral#1}}
\newcommand{\ronum}[1]{\expandafter{\romannumeral#1}}
\DeclareRobustCommand{\LaTeXZ}{%
  \LaTeX\kern-.05em4\kern-.1em
  {\raisebox{-0.2ex}{$\scriptstyle\text{ZEUS}$}}\xspace}

\DeclareMathAlphabet{\mathbf}{OT1}{cmr}{bx}{sl}
\newcommand{\eVdist}{\kern-0.06667em}

\newcommand{\Gev}{{\text{Ge}\eVdist\text{V\/}}}

\newcommand{\gev}{{\,\text{Ge}\eVdist\text{V\/}}}

\newcommand{\Tesla}{\,\text{T}}

\newcommand{\slashfrac}[2]{%
  \raisebox{0.5ex}{\ensuremath #1}\kern-0.12em/\kern-0.08em
  \raisebox{-.8ex}{\ensuremath #2}}

\newcommand{\sqr}[3]{%
    {\vcenter{\hrule height.#3ex\hbox{\vrule width.#2ex height#1ex
     \kern#1ex\vrule width.#3ex}\hrule height.#2ex}}}

\newcommand{\widebar}[1]{%
   \mkern1.5mu\overline{\mkern-1.5mu#1\mkern-1.mu}\mkern1.mu}
\catcode`\@=11 
\newcommand{\parenbar}{\mathpalette\p@renb@r}
\def\p@renb@r#1#2{\vbox{%
  \ifx#1\scriptscriptstyle \dimen@.7em\dimen@ii.2em\else
  \ifx#1\scriptstyle \dimen@.8em\dimen@ii.25em\else
  \dimen@1em\dimen@ii.4em\fi\fi \offinterlineskip
  \ialign{\hfill##\hfill\cr
    \vbox{\hrule width\dimen@ii}\cr
    \noalign{\vskip-.3ex}%
    \hbox to\dimen@{$\mathchar300\hfil\mathchar301$}\cr
    \noalign{\vskip-.3ex}%
    $#1#2$\cr}}}
\catcode`\@=12 

\newcommand{\ppbar}{\parenbar{p}}

\newcommand{\pbar}{\widebar{p}}

\newcommand{\qbar}{\widebar{q}}

\newcommand{\cbar}{\widebar{c}}
\newcommand{\bbar}{\widebar{b}}

\newcommand{\IP}{{\rm I$\kern-0.01667em$P}\xspace}

\mathchardef\qsm=63
\mathchardef\pls=43
\mathchardef\mns=512
\mathchardef\plm=518
\mathchardef\eql=61
\mathchardef\smallleft=300
\mathchardef\smallright=301
\mathchardef\les=316
\mathchardef\gre=318
\mathchardef\leq=532
\mathchardef\grq=533

\catcode`\@=11 
\newcounter{pict@width}
\newcounter{pict@height}
\newlength{\pict@scale}
\newcommand{\psfigadd}[4]{%
\setcounter{pict@width}{1*\ratio{#2+\pict@scale/2}{\pict@scale}}
\setcounter{pict@height}{1*\ratio{#3+\pict@scale/2}{\pict@scale}}
\setlength{\unitlength}{\pict@scale}
\hbox to #2{\hspace{-\fill}\begin{picture}(\thepict@width,\thepict@height)
\put(0,0){\psfig{figure=#1,width=#2,height=#3,clip=}}
\SetScale{0.283466457}
\SetWidth{1.763889}
{#4}
\end{picture}}
}
\newcounter{pict@widthfst}
\newcounter{pict@widthscd}
\newcounter{pict@widthtot}
\newcommand{\psfigaddtwo}[7]{%
\setcounter{pict@widthfst}{1*\ratio{#2+\pict@scale/2}{\pict@scale}}
\setcounter{pict@widthscd}{1*\ratio{#2+#4+\pict@scale/2}{\pict@scale}}
\setcounter{pict@widthtot}{1*\ratio{#2+#4+#6+\pict@scale/2}{\pict@scale}}
\setcounter{pict@height}{1*\ratio{#3+\pict@scale/2}{\pict@scale}}
\setlength{\unitlength}{\pict@scale}
\hbox{\hspace{-\fill}\begin{picture}(\thepict@widthtot,\thepict@height)
\put(0,0){\psfig{figure=#1,width=#2,height=#3,clip=}}
\put(\thepict@widthscd,0){\psfig{figure=#5,width=#6,height=#3,clip=}}
\SetScale{0.283466457}
\SetWidth{1.763889}
{#7}
\end{picture}}
}
\newcommand{\psfigror}[4]{%
\setcounter{pict@width}{1*\ratio{#2+\pict@scale/2}{\pict@scale}}
\setcounter{pict@height}{1*\ratio{#3+\pict@scale/2}{\pict@scale}}
\setlength{\unitlength}{\pict@scale}
\hbox{\begin{picture}(\thepict@width,\thepict@height)
\put(0,\thepict@height){\psfig{figure=#1,width=#3,height=#2,clip=,angle=270}}
\SetScale{0.283466457}
\SetWidth{1.763889}
{#4}
\end{picture}}
}
\newcommand{\psfigrol}[4]{%
\setcounter{pict@width}{1*\ratio{#2+\pict@scale/2}{\pict@scale}}
\setcounter{pict@height}{1*\ratio{#3+\pict@scale/2}{\pict@scale}}
\setlength{\unitlength}{\pict@scale}
\hbox{\begin{picture}(\thepict@width,\thepict@height)
\put(0,0){\psfig{figure=#1,width=#3,height=#2,clip=,angle=90}}
\SetScale{0.283466457}
\SetWidth{1.763889}
{#4}
\end{picture}}
}
\catcode`\@=12 
\newlength\listtextwidth

\catcode`\@=11 
\newlength{\@tabfninsert}
\newlength{\@tabfnwidth}
\newcommand{\tabfootnote}[2]{%
  \setlength{\@tabfninsert}{0.8em}
  \setlength{\@tabfnwidth}{\textwidth}
  \addtolength{\@tabfnwidth}{-\@tabfninsert}
  \addtolength{\@tabfnwidth}{-0.4em}
  \noindent\makebox[\@tabfninsert][r]{\footnotesize$^{#1}$\hfil}\hfill%
  \parbox[t]{\@tabfnwidth}{\footnotesize #2\hfill}}
\catcode`\@=12 

\newcommand{\ptr}{\mbox{$p_{\rm T}^{\rm rel}$}}

\def\citeCTD{{\cite{%
nim:a279:290,*npps:b32:181,*nim:a338:254%
}}\xspace}
\def\citeCAL{{\cite{%
nim:a309:77,*nim:a309:101,*nim:a321:356,*nim:a336:23%
}}\xspace}

\begin{document}
\title{Beauty photoproduction measured using decays into muons
in dijet events  in $ep$ collisions at $\sqrt{s}=318\gev$}
\author{ZEUS Collaboration}
\draftversion{4.0}
\date{December 2003\\ Revised September 2006}
\prepnum{DESY-03-212}
\abstract{The photoproduction of beauty quarks 
 in events with two jets
and a muon
has been measured with the ZEUS detector at HERA
using an integrated luminosity of 110~pb$^{- 1}$.
The fraction of jets containing $b$ quarks
was extracted from the transverse momentum distribution of the muon
relative to the closest jet.
Differential cross sections
for beauty production as a function of the  transverse
momentum and pseudorapidity of the muon, of the 
associated jet and of $x_{\gamma}^{\rm jets}$,
the fraction of the photon's momentum participating in the hard process,
 are compared with MC models
and  QCD predictions made at next-to-leading order.
The latter give a good description of the data.}

\makezeustitle

\def\3{\ss}                                                                                        
\pagenumbering{Roman}                                                                              
                                                   %
\begin{center}                                                                                     
{                      \Large  The ZEUS Collaboration              }                               
\end{center}                                                                                       
  S.~Chekanov,                                                                                     
  M.~Derrick,                                                                                      
  D.~Krakauer,                                                                                     
  J.H.~Loizides$^{   1}$,                                                                          
  S.~Magill,                                                                                       
  S.~Miglioranzi$^{   1}$,                                                                         
  B.~Musgrave,                                                                                     
  J.~Repond,                                                                                       
  R.~Yoshida\\                                                                                     
 {\it Argonne National Laboratory, Argonne, Illinois 60439-4815}, USA~$^{n}$                       
\par \filbreak                                                                                     
  M.C.K.~Mattingly \\                                                                              
 {\it Andrews University, Berrien Springs, Michigan 49104-0380}, USA
% ...
\newpage                                                                                           
$^{\    1}$ also affiliated with University College London, London, UK \\                          
% ...
                                                           %
                                                           %
\newpage   
                                                           %
                                                           %
\begin{tabular}[h]{rp{14cm}}                                                                       
$^{a}$ &  supported by the Natural Sciences and Engineering Research                               
          Council of Canada (NSERC) \\                                                             
$^{b}$ &  supported by the German Federal Ministry for Education and                               
          Research (BMBF), under contract numbers HZ1GUA 2, HZ1GUB 0, HZ1PDA 5, HZ1VFA 5\\         
$^{c}$ &  supported by the MINERVA Gesellschaft f\"ur Forschung GmbH, the                          
          Israel Science Foundation, the U.S.-Israel Binational Science                            
          Foundation and the Benozyio Center                                                       
          for High Energy Physics\\                                                                
$^{d}$ &  supported by the German-Israeli Foundation and the Israel Science                        
          Foundation\\                                                                             
$^{e}$ &  supported by the Italian National Institute for Nuclear Physics (INFN) \\                
$^{f}$ &  supported by the Japanese Ministry of Education, Culture,                                
          Sports, Science and Technology (MEXT) and its grants for                                 
          Scientific Research\\                                                                    
$^{g}$ &  supported by the Korean Ministry of Education and Korea Science                          
          and Engineering Foundation\\                                                             
$^{h}$ &  supported by the Netherlands Foundation for Research on Matter (FOM)\\                   
$^{i}$ &  supported by the Polish State Committee for Scientific Research,                         
          grant no. 620/E-77/SPB/DESY/P-03/DZ 117/2003-2005\\                                      
$^{j}$ &  partially supported by the German Federal Ministry for Education                         
          and Research (BMBF)\\                                                                    
$^{k}$ &  partly supported by the Russian Ministry of Industry, Science                            
          and Technology through its grant for Scientific Research on High                         
          Energy Physics\\                                                                         
$^{l}$ &  supported by the Spanish Ministry of Education and Science                               
          through funds provided by CICYT\\                                                        
$^{m}$ &  supported by the Particle Physics and Astronomy Research Council, UK\\                   
$^{n}$ &  supported by the US Department of Energy\\                                               
$^{o}$ &  supported by the US National Science Foundation\\                                        
$^{p}$ &  supported by the Polish State Committee for Scientific Research,                         
          grant no. 112/E-356/SPUB/DESY/P-03/DZ 116/2003-2005,2 P03B 13922\\                       
$^{q}$ &  supported by the Polish State Committee for Scientific Research,                         
          grant no. 115/E-343/SPUB-M/DESY/P-03/DZ 121/2001-2002, 2 P03B 07022\\                    
\end{tabular}                                                                                      
                                                           %
                                                           %

\pagenumbering{arabic} 
\pagestyle{plain}
\section{Introduction}
\label{sec-int}

The production of beauty quarks in $ep$ collisions at HERA
is a stringent test for perturbative Quantum Chromodynamics (QCD)
 since the large 
$b$-quark mass ($m_b \sim 5 \gev$) provides a hard scale that should
ensure reliable predictions. When $Q^2$, the negative squared four-momentum 
exchanged at the electron vertex, is small the reaction 
$ep \rightarrow e' b \bbar X$ can be considered as a photoproduction process
 in which  a quasi-real photon, emitted by the incoming electron, interacts 
with the proton.

For $b$-quark transverse momenta comparable to the $b$-quark mass,  
next-to-leading-order (NLO) QCD calculations in which the $b$ quark is 
generated dynamically are expected to provide accurate predictions for $b$ 
photoproduction \cite{frixione2,*frixione3}. The corresponding leading-order (LO) QCD processes are   
the direct-photon process, in which the quasi-real photon enters directly in 
the hard interaction,
 $\gamma g \rightarrow b\bbar$, and the resolved-photon process, 
in which the photon acts as a source of partons
that take part in the hard interaction ($g g \rightarrow b \bbar$ or $q \qbar \rightarrow b \bbar$).
 
The  beauty-production cross section has been measured
 in $p\ppbar$ collisions at the ISR \cite{ISR}, S$p\pbar$S
 \cite{beautyUA1,*beautyUA1b} and
Tevatron colliders \cite{beautyCDF1,*beautyCDF2,*beautyCDF3,*beautyCDF4,*beautyCDF5,*beautyD00,*beautyD01,*beautyD02,*beautyD03},
in $\gamma \gamma$  interactions at LEP \cite{beautyLEP1} and in fixed-target $\pi N$ \cite{WA78,*E706} and $pN$  \cite{E771,*E789,*HERAB} experiments.
Apart from the S$p\pbar$S data and the fixed-target experiments,
the results were significantly above the NLO QCD prediction.
The H1 measurement in $ep$ interactions at HERA
\cite{h1bmuon1} found a cross section
significantly larger than the prediction.
The previous ZEUS measurement \cite{ZEUSelec} 
was above, but consistent with, the prediction.

This paper reports a measurement of beauty photoproduction
in events with two jets and a muon, 
$ep \rightarrow e' \;  b \bbar \; X \rightarrow e' \; jj \mu \; X'$,
for $Q^2<1 \gev^2$.
\section{Experimental set-up}
\label{sec-exp}
The data sample used in this analysis  corresponds to an integrated luminosity 
${\cal L}=110.4 \pm 2.2 ~\rm{pb}^{-1}$, collected by the ZEUS detector
in the years 1996-1997 and 1999-2000.
During the 1996-97 data taking,
HERA provided collisions between an electron\footnote{Electrons and positrons are not distinguished
in this paper and are both referred to as electrons.}
beam of  $E_e=27.5 \gev$ and a proton beam of
$E_p=820\gev$, corresponding to a  centre-of-mass energy  $\sqrt s=300\gev$
(${{\cal L}_{300}}=38.0\pm 0.6~ \rm{pb}^{-1}$). In the years 1999-2000,
the proton-beam energy was 
$E_p=920\gev$,  corresponding to $\sqrt s=318\gev$
(${{\cal L}_{318}}=72.4\pm 1.6~\rm{pb}^{-1}$).

A detailed description of the ZEUS detector can be found
elsewhere~\cite{zeus:1993:bluebook}. A brief outline of the
components that are most relevant for this analysis is given
below.

Charged particles are tracked in the central tracking detector (CTD)~\citeCTD,
which operates in a magnetic field of $1.43\Tesla$ provided by a thin
superconducting solenoid. The CTD consists of 72~cylindrical drift chamber
layers, organized in nine superlayers covering the polar-angle\footnote{The ZEUS coordinate system is a right-handed Cartesian system, with the $Z$ axis pointing in the proton beam direction, referred to as the ``forward direction'', and the $X$ axis pointing left towards the centre of HERA. The coordinate origin is at the nominal interaction point.}
 region
\mbox{$15^\circ<\theta<164^\circ$}. The transverse-momentum resolution for
full-length tracks is $\sigma(p_{T})/p_{T}=0.0058p_{T}\oplus0.0065\oplus0.0014/p_{T}$, with $p_{T}$ in $\Gev$.

The high-resolution uranium-scintillator calorimeter (CAL)~\citeCAL consists
of three parts: the forward (FCAL), the barrel (BCAL) and the rear (RCAL)
calorimeters. Each part is subdivided transversely into towers and
longitudinally into one electromagnetic section (EMC) and either one (in RCAL)
or two (in BCAL and FCAL) hadronic sections (HAC). The smallest subdivision of
the calorimeter is called a cell.  The CAL energy resolutions,
as measured under test-beam conditions, are $\sigma(E)/E=0.18/\sqrt{E}$ for
electrons and $\sigma(E)/E=0.35/\sqrt{E}$ for hadrons, with $E$ in $\Gev$.

The muon system consists of rear, barrel (R/BMUON) \cite{brmuon} and forward (FMUON) \cite{zeus:1993:bluebook} tracking detectors.
The B/RMUON consists of limited-streamer (LS) tube chambers
placed behind the BCAL (RCAL), inside and outside a magnetized iron yoke
surrounding the CAL.
The barrel and rear muon chambers cover polar angles from $34^{\rm o}$
to $135^{\rm o}$ and from $135^{\rm o}$ to $171^{\rm o}$, respectively.
The FMUON consists of six trigger planes of LS tubes
and four planes of drift chambers
covering the angular region from $5^{\rm o}$ to $32^{\rm o}$.
The muon system exploits the magnetic field of the iron yoke and,
in the forward direction, of two iron toroids magnetized to $\sim 1.6$~T
to provide an independent measurement of the muon momentum.

The luminosity was measured using the bremsstrahlung process $ep \to e p \gamma$.
The resulting small-angle energetic photons were measured by the luminosity
monitor~\cite{Desy-92-066,*zfp:c63:391,*acpp:b32:2025}, a lead-scintillator calorimeter placed in the HERA tunnel at $Z = -107$ m.

\section{Data Selection}

The data were selected online by requiring either
a high-momentum muon reaching the external B/RMUON chambers
or  two jets reconstructed in the CAL.
A dedicated trigger requiring two jets and a muon
with looser jet and muon thresholds was also used in the last
part of the data taking.

Muons were reconstructed offline using the following procedure: a
muon track was found in the inner and outer B/RMUON chambers 
or crossing at least 4 FMUON 
planes, then a match in position and momentum to a CTD track was required.
The angular acceptance of the F/B/RMUON and of the CTD, and the requirement that the muons
reach the external chambers define three regions of good acceptance:
\begin{eqnarray}
\label{e:mucuts}
 \textrm{rear   } & -1.6<\eta^{\mu}<-0.9, &p^{\mu}>2.5\gev; \nonumber \\
 \textrm{barrel } & -0.9<\eta^{\mu}<1.3,  &p_{T}^{\mu}>2.5\gev;\\
 \textrm{forward} &  1.48<\eta^{\mu}<2.3,   &p^{\mu}>4\gev,~p_{T}^{\mu}>1\gev; \nonumber
\end{eqnarray}
where $\eta^{\mu}$, $p^{\mu}$ and $p_{T}^{\mu}$ are the muon pseudorapidity,
momentum and transverse momentum, respectively.

The hadronic system (including the muon) was reconstructed from energy-flow 
objects (EFOs)~\cite{ZEUSefo,*briskin} 
which combine the information from calorimetry and tracking, corrected
for energy loss in dead material.
A reconstructed four-momentum $(p_X^i,p_Y^i,p_Z^i,E^i)$ was assigned
to each EFO.

Jets were reconstructed from EFOs 
using the $k_{T}$ algorithm \cite{ktclus} in the
longitudinally invariant mode  \cite{ktalgn}.
The  $E$ recombination scheme, which produces massive jets whose
four-momenta are the sum of the four-momenta of the clustered objects,
was used. Muons were associated with jets by the  $k_{T}$ algorithm:
 if the EFO corresponding to a reconstructed muon was included in a
jet then the muon was considered to be associated with the jet. 

The event inelasticity $y$ was reconstructed from the
Jacquet-Blondel estimator
$y_{\rm JB}=(E-p_Z)/(2E_e)$ \cite{JacquetBlondel}, 
where \mbox{$E-p_Z=\sum_i E^i-p_Z^i$}
and the sum runs over all EFOs.

A sample of events with one muon and two jets was selected by requiring:
\begin{itemize}
\item $\ge 1$ muon in one of the three muon-chamber 
      regions defined in Eq.~(\ref{e:mucuts});
\item $\ge 2$ jets with pseudorapidity $|\eta^{\rm \: jet}| < 2.5$,
      and transverse momentum $p_{T}^{\rm \: jet}>7 \gev$
for the highest-$p^{\rm \: jet}_{T}$ jet and $p_{T}^{\rm \: jet}>6 \gev$ 
for the second-highest-$p^{\rm \:jet}_{T}$ jet;
\item that the muon was associated with any jet
 with $p_{T}^{\rm \: jet}>6 \gev$ and 
$|\eta^{\: {\rm jet}}|<2.5$. To assure a reliable $\ptr$
measurement (see Section~\ref{s:signal}),
the residual jet transverse momentum, calculated excluding the associated 
muon, was required to be greater than $2  \gev$;
\item a reconstructed vertex compatible with the nominal interaction point;
\item no scattered-electron candidate found in the CAL;
\item $0.2<y_{\rm JB}<0.8$.
\end{itemize}
The last two cuts suppress  background from high-$Q^2$ events and 
from non-$ep$ interactions, and correspond to an effective cut  $Q^2<1\gev^2$
and $0.2<y<0.8$.

After this selection, a sample of 3660 events remained.

\section{Acceptance corrections and background simulation}

To evaluate the detector acceptance and to provide the signal and background
 distributions, Monte Carlo (MC) samples of beauty, charm and light-flavour
 (LF) events were generated,
corresponding respectively to six, five and three times the luminosity 
of the data. 
{\sc Pythia 6.2} \cite{pythia62,pythia61exc}
was used as the reference MC, 
and {\sc Herwig 6.1} \cite{herwig} for systematic checks. 
The branching ratios for direct semi-leptonic $b \rightarrow \mu X$ decays
and for indirect cascade decays into muons
via charm, anti-charm, $\tau^\pm$ and $J/\psi$,
were set to 
 ${\cal B}_{\rm dir}=0.106\pm0.002$ and
 ${\cal B}_{\rm indir}=0.103\pm0.007$ \cite{pdg}, respectively.
The distribution of the decay-lepton momentum in
the B-meson centre-of-mass system
from {\sc Pythia} and {\sc Herwig} has been compared with
measurements from $e^+e^-$ collisions \cite{belle,*babar} and found
to be in good agreement.
The generated events were passed through a full simulation of the ZEUS
detector based on GEANT 3.13~\cite{geant}.
 They were then subjected to the same trigger
requirements and processed by the same reconstruction programs as for the
data.

\begin{figure}[t]
\centerline{{ \leavevmode \epsfxsize=15cm \epsffile{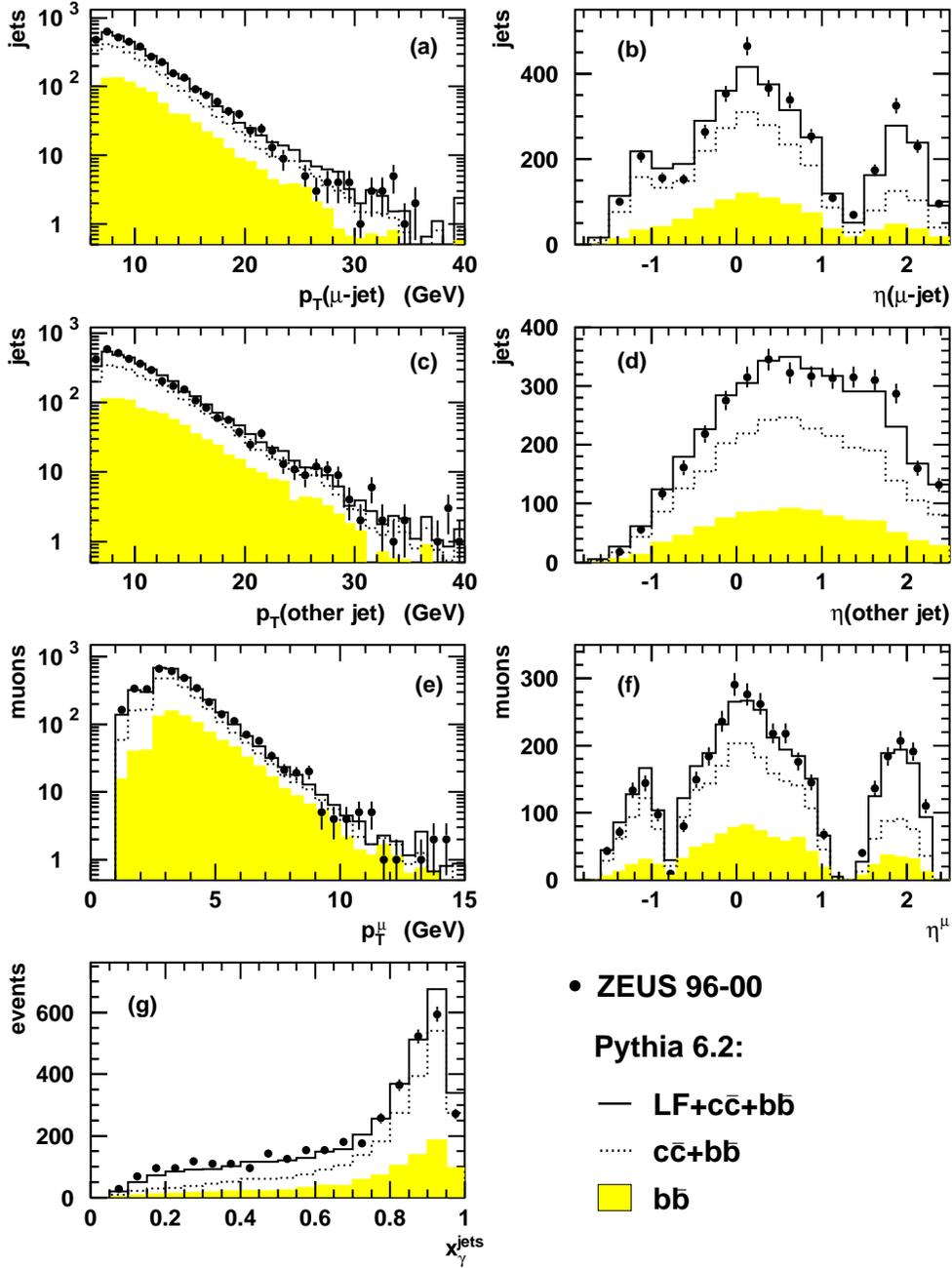}}}
\caption[this space for rent]{Distributions
 for the dijet-plus-muon sample (points)
 compared to the predictions of the {\sc Pythia} Monte Carlo (full line)
 normalized to the data.
 The shaded histogram shows the beauty component and the dotted line is the
 sum  of charm and beauty. 
 The plots show (a) the transverse momentum of the jet associated 
 with the muon; (b) its pseudorapidity;
 (c) the transverse momentum of the highest-$p_T$  non-$\mu$-associated jet;
 (d) its pseudorapidity; (e) the transverse momentum of the muon; 
 (f) its pseudorapidity; (g) the distribution of $x_\gamma^{\rm jets}$.}
\label{f:f1}
\end{figure}

Figure \ref{f:f1} shows the kinematic distributions of
$p_{T}^{\rm \: jet}$ and $\eta^{\rm \: jet}$ for the jet associated
with the muon, as well as for highest-$p_T$ jet
that was not associated with a muon (other jet). The muon kinematic variables 
$p_{T}^{\mu}$ and $\eta^{\mu}$ are displayed, as well as 
$x_\gamma^{\rm jets}$, the fraction of the total hadronic
$E-p_Z$ carried by the two highest-$p_{T}$ jets\footnote{
$x_{\gamma}^{\rm jets}$ is the massive-jets analogue of the 
$x_{\gamma}^{\rm obs}$ variable used for massless jets
 in other ZEUS publications\cite{ZEUSxg}.}

\begin{equation}
x_{\gamma}^{\rm jets}=\frac{\sum_{j=1,2} (E^{\rm jet}-p^{\rm jet}_Z)}{E-p_Z}.
\end{equation}

The data are compared in shape to the {\sc Pythia} MC sample
in which the relative
fractions of beauty, charm and LF were mixed according to the cross 
sections predicted by
the simulation. The comparison shows that the main
features of the dijet-plus-muon sample are well reproduced by this MC 
mixture.
The {\sc Pythia} MC predicts that the non-$b\bbar$ background 
comprises
57\% prompt muons from charmed-hadron decays
and  43\% muons from light-flavour hadrons,
mostly due to in-flight decays of
$\pi$ and $K$ mesons, with a small amount ($\sim 5\%$ of the LF component)
from muons produced in interactions with the detector material.
The punch-through  contribution is negligible.
The {\sc Herwig} Monte Carlo (not shown) also gives a good
 description of the data.

The detector acceptance for the final cross sections was 
calculated using the $b\bbar$
{\sc Pythia} Monte Carlo,
in which events were reweighted such that the transverse momentum distribution
of the $b$ quark agreed with that of the NLO calculations. The effects of this
reweighting on the distributions in Fig.\ref{f:f1} was small.

\section{Signal extraction and cross section measurement}
\label{s:signal}
Because of the large $b$-quark mass,
muons from semi-leptonic B-hadron decays
tend to be produced with large transverse momentum
with respect to the direction of the jet containing the B-hadron.
The beauty signal was  extracted by
exploiting the distribution of the transverse momentum of the muon
with respect to the momentum of the
rest of the associated jet, \ptr, defined as:
\begin{equation}
\ptr=\frac{ | \mathbf{p}^{\:\mu} \times ( \mathbf{p}^{\rm \: jet}-\mathbf{p}^{\:\mu} ) |}
          { | \mathbf{p}^{\rm \: jet}-\mathbf{p}^{\:\mu} | },
\end{equation}
where $\mathbf{p}^{\:\mu}$ is the muon and $\mathbf{p}^{\rm \: jet}$
the jet momentum vector.
Figure 2a shows the 
 distributions, normalized to unity,
of the reconstructed muon $\ptr$ as obtained from the {\sc Pythia} MC,
for beauty ($f_{\mu}^{b\bbar,{\rm MC}}$), charm
  ($f_{\mu}^{c\cbar,{\rm MC}}$) and LF ($f_{\mu}^{\rm LF,MC}$) events.
The $\ptr$ distribution for beauty peaks
at $\sim 2$~GeV and is well separated from those from charm
and LF which are peaked close to zero.
Since the shapes of
$f_{\mu}^{\rm LF,MC}$ and $f_{\mu}^{c\cbar,{\rm MC}}$
 are very similar, 
the fraction of beauty ($a_{b\bbar}$)
and background ($a_{\rm bkg}$) events in the sample was
obtained from a two-component fit to the shape of the 
measured $\ptr$ distribution
 $f_{\mu}$  
with a beauty and a background component:
\begin{equation}
\label{e:ptrfit}
f_{\mu}=a_{\rm bkg} f_{\mu}^{\rm bkg}
  + a_{b\bbar}  f_{\mu}^{b\bbar},
\end{equation}
where the $\ptr$ distribution of beauty, $f_{\mu}^{b\bbar}$, 
was taken from the {\sc Pythia} MC: 
$f_{\mu}^{b\bbar}=f_{\mu}^{b\bbar,{\rm MC}}$, and that of the background,
$f_{\mu}^{\rm bkg}$, was obtained as explained below.

The distribution $f_{\mu}^{\rm bkg}$
was obtained from the sum of a LF, $f_{\mu}^{\rm LF}$, and a charm, 
$f_{\mu}^{c\cbar}$, distribution
weighted according to the charm fraction $r$ obtained
from  the charm and LF cross sections given by {\sc Pythia}:
\begin{equation}
f_{\mu}^{\rm bkg} = r  f_{\mu}^{c\cbar} + (1-r)  f_{\mu}^{\rm LF}.
\end{equation}

The distribution $f_{\mu}^{\rm LF}$ 
can be obtained  from the $\ptr$ distribution of a sample of
CTD tracks not identified as muons but
fulfilling the same momentum and angular requirements 
applied to muons (called ``unidentified tracks'' in the following).
The $\ptr$ distribution for unidentified tracks,   $f_x$,
is expected to be similar to $f_{\mu}^{\rm LF}$,
under the assumption that the probability for an unidentified
track (typically a $\pi$ or a $K$ meson) to be identified as a muon,
$P_{x\rightarrow \mu}$, does not depend strongly on $\ptr$.
This assumption is validated by the MC, since the MC distributions
for the LF background, $f_{\mu}^{\rm LF,MC}$, and for the
unidentified tracks, $f^{\rm MC}_x$, are indeed very similar,
as shown in Fig. 2a.

Figure 2b shows the $f_x$ distribution  obtained  from
a dijet sample selected without muon requirements.
The shape obtained from {\sc Pythia}, $f_x^{\rm MC}$,
underestimates the tail (for example by $24\%$ at 2.625 GeV).
The $\ptr$ shape of the LF background was therefore obtained as
\begin{equation}
 f_{\mu}^{\rm LF}= f_x \frac{f_{\mu}^{\rm LF, MC}}{f_x^{\rm MC}},
\end{equation} 
where the ratio $f_{\mu}^{\rm LF, MC}/f_x^{\rm MC}$ 
is a MC-based correction that accounts for  
possible differences between 
$f_{\mu}^{\rm LF}$ and $f_x$ due to a residual 
$\ptr$ dependence of $P_{x\rightarrow \mu}$ and to  
 the charm and beauty contamination
($\sim 28\%$ and $\sim 2\%$ of respectively)
in the dijet sample.
 
The data cannot be used to extract
the distribution $f_{\mu}^{c\cbar}$. Two cases
were therefore considered:
 the distribution given
by the {\sc Pythia} MC, $f_{\mu}^{c\cbar,{\rm MC}}$, and the
distribution obtained from the unidentified track sample, as in the case
of the LF background: $f_x\frac{f_{\mu}^{c\cbar,{\rm MC}}}{f_x^{\rm MC}}$.
The average  of these two cases 
was then taken as the nominal $f_{\mu}^{c\cbar}$:

Figure \ref{f:fit} shows the result of the $\ptr$ fit 
for muons in the rear, barrel and forward regions.
The sum of the two components 
reproduces the data reasonably well.
The fraction of $b$ in the total sample of dijet events with a muon
is $a_{b\bbar}= 0.224 \pm 0.017 \textrm{ (stat.)}$. In the determination
 of the cross sections, the fraction of beauty events
in the data was extracted by a fit performed in each cross-section bin.

All the cross sections reported in Section \ref{results},
with the exception of that for $b$ quarks,
are inclusive muon (or $b$-jet) cross sections,
obtained by counting muons (or $b$-jets) rather than events.
Muons coming from both direct and indirect
$b$ decays are considered to be part of the signal.  
The cross sections are
given for dijet events passing
the following requirements:
$Q^2<1 \gev^2$, $0.2<y<0.8$ and
at least two hadron-level jets with 
$p_{T}^{\rm \: jet_1}>7 \gev$, $p_{T}^{\rm \: jet_2}>6 \gev$
and $\eta^{\rm \: jet_1},\eta^{\rm \: jet_2}<2.5$.
These jets were defined using
the $k_{T}$ algorithm on stable hadrons,
where the weakly decaying B hadrons were 
considered stable. For dijet events with a muon passing the cuts of Eq.~(\ref{e:mucuts}), the acceptance varies from 10\% at low $p_{T}^{\mu}$ to
20\% at large  $p_{T}^{\mu}$.

The cross sections were measured 
from data collected at two different centre-of-mass energies,
$\sqrt{s}\!=\!300 \gev$ and
 $\sqrt{s}\!=\! 318 \gev$. They were corrected
to $\sqrt{s}\! =\! 318 \gev$ using the NLO QCD prediction.
The effect of this correction on the final cross section is $\sim 2\%$.

\section{Theoretical predictions and uncertainties}

The measured cross sections are compared to NLO QCD predictions based on the 
FMNR \cite{fmnr} program.
The parton distribution functions used for the nominal prediction were
GRVG-HO \cite{GRVLO} for the photon and CTEQ5M \cite{CTEQ5M} for the proton.
 The $b$-quark mass was set
to $m_{b}=4.75~\gev$, and the renormalisation and factorisation scales 
to the transverse mass, 
$\mu_{r}\!=\!\mu_{f}\!=m_{T}\!=\!\sqrt{\frac{1}{2}\left( (p_{T}^{b})^2+(p_{T}^{\bbar})^2 \right)+m_{b}^{2}}$,
where $p_{T}^{b(\bbar)}$ is the transverse momentum
of the $b$ ($\bbar$) quark in the laboratory frame.
Jets were reconstructed by running the $k_{T}$ algorithm on the four
momenta of the $b$ and $\bbar$ quarks and of the third light parton
(if present) generated by the program.
The fragmentation of the $b$ quark into a B hadron was simulated by rescaling
 the quark three-momentum
(in the frame in which   $p^b_Z=-p^{\bbar}_Z$, obtained with a boost along 
$Z$) 
according to the Peterson \cite{peterson} fragmentation function with $\epsilon = 0.0035$.
The muon momentum  was generated isotropically in the B-hadron rest frame
 from the decay spectrum given by {\sc Pythia}, which
 is in good agreement with measurements made at B factories
 \cite{belle,*babar}.

To evaluate the uncertainty on the NLO calculations, the $b$-quark mass
and  the renormalisation and factorisation scales were varied simultaneously,
to maximize the change,
from $m_b=4.5~\rm{GeV}$ and  $\mu_{r}\!=\!\mu_{f}\!=m_{T}/2$ to
$m_{b}=5.0~\rm{GeV}$ and $\mu_r\!=\!\mu_f=2m_{T}$, producing a variation
in the cross section from $+34\%$ to $-22\%$.
The effect on the cross section of a variation of the Peterson parameter $\epsilon$ from 0.002 to 0.0055\cite{NasonOleari} and of a change of the fragmentation function from the Peterson to the Kartvelishvili parametrisation (with  $\alpha = 13$, as obtained from comparisons between NLO QCD and MC distributions)
\cite{Kartvelishvili,CacciariNason} was less than $\pm 3\%$. 
The effects of using different sets of parton densities and of
a variation of the strong coupling constant
($\Lambda^{(5)}_{\rm QCD}=0.226\pm0.025~{\rm MeV}$) were all within $\pm 4\%$. 

The NLO cross sections, calculated for jets made of partons,
were corrected for jet hadronization effects to allow a direct comparison
 with the measured hadron-level cross sections.
The corrections were derived from the MC simulation as the ratio of the
hadron-level to the parton-level MC cross section,
where the parton level is defined as
being the result of the parton showering stage of the simulation.
The average between the corrections obtained from 
{\sc Pythia} and {\sc Herwig} was taken as the central value
and their difference as the uncertainty.
The effect of the hadronization correction was largest in the rear region,
where the cross section was reduced 
by $(20 \pm 6)\%$ and smallest at large 
$p^{\mu}_{T}$ where
it was reduced  by $(3.0 \pm 0.3)\%$.

The measured cross sections are also compared to the absolute
predictions of two MC models, {\sc Pythia 6.2} and {\sc Cascade 1.1}.
The predictions of {\sc Pythia 6.2} were obtained \cite{pythia61exc} by
mixing direct- ($\gamma g \rightarrow b \bbar$) and resolved-photon
 ($g g \rightarrow b \bbar$, $q \qbar \rightarrow b \bbar$ )
flavour-creation processes calculated using massive matrix elements
 and the flavour-excitation (FE) processes
 ($b g  \rightarrow b g$,  $b q  \rightarrow b q$),
in which a heavy quark is extracted from the photon or proton parton density.
The FE processes contribute about $27\%$ of the total
$b\bbar$ cross section.
The small ($\sim 5\%$) contribution from gluon splitting in parton showers
($g  \rightarrow b \bbar$) was not included.
The parton density CTEQ4L \cite{CTEQLO} was used for the proton and
GRVG-LO \cite{GRVLO} for the
photon; the $b$-quark mass was set to $4.75$ GeV and the $b$-quark string
fragmentation was performed according to the Peterson function with $\epsilon=0.0041$ \cite{opal}.

{\sc Cascade} \cite{cascade1,*cascade2} is a Monte Carlo implementation 
of the CCFM evolution equations \cite{CCFM1,*CCFM2}.
Heavy-quark production is obtained from the  $O(\alpha_s)$ matrix elements for
the process $\gamma g^* \rightarrow b\bbar $, in which the initial 
gluon can be off-shell.
The  gluon density, unintegrated
in transverse momentum $(k_{T})$, is
 obtained from an analysis of the proton structure functions  based on
 the CCFM equations \cite{ccfmfits};
 in the event generation the gluon density used corresponds to the set 
 named {``J2003 set 2''}. The mass of the $b$ quark
 was set to 4.75 GeV and $\alpha_s$ was evaluated at the scale $m_{T}$.
 As for {\sc Pythia}, the $b$-quark string fragmentation was performed 
 according to the Peterson function with $\epsilon = 0.0041$.

\section{Systematic uncertainties}
The main experimental uncertainties are described below:
\begin{itemize}
\item the muon acceptance, including the efficiency of the muon chambers,
      of the reconstruction
      and of the MUON-CTD matching, 
      is known to about $10\%$ from a study based on an independent 
      dimuon  sample \cite{tesiMonica};
\item the uncertainty on the $\ptr$ shape of the LF and charm background
      was evaluated by:
\begin{itemize} 
  \item varying the charm fraction in the background, $r$, by
        $\pm 20\%$. This range was obtained by fixing the absolute
        charm-MC normalisation to a measurement of the
        charm dijet cross section \cite{ZEUScharm} and using
        the {\sc Pythia} or {\sc Herwig} MC to extrapolate to the 
        kinematic range of the present measurement;
  \item varying the jet-track association in the unidentified-track sample;
  \item  extracting $f_{\mu}^{\rm LF}$ from
         a sample of unidentified CTD tracks, reweighted 
         with a MC-based parametrisation of $P_{x\rightarrow \mu}$
         depending on polar angle and momentum;
  \item varying the $\ptr$ shape of the charm component of the background
        between the prediction from {\sc Pythia} and the value
        obtained from the unidentified track sample;
  \item using {\sc Herwig} instead of {\sc Pythia} to simulate the background.
 \end{itemize}
      The total uncertainty from these sources is about $10\%$.
      As a cross-check, a different definition of $\ptr$ was used to extract 
      the beauty fraction, namely the transverse momentum of the muon with
      respect to the whole jet, including the muon itself,
      as used in a previous ZEUS publication \cite{ZEUSelec}.
      The results were found to be in good agreement;
\item the $2\%$ uncertainty on the direct-decay branching ratio 
       ${\cal B}_{\rm dir}$
      introduces a $2\%$ uncertainty on the $b$-jet and on the $b$-quark
      cross sections while it has no effect on the visible muon cross sections. 
      The $7\%$ uncertainty on the branching ratio for indirect decays 
      ${\cal B}_{\rm indir}$
      produces an uncertainty of 1\% on the measured cross sections;
\item the uncertainties on the dijet selection, 
      on the energy scale, on the jet and $y_{\rm JB}$ resolution and  
      trigger efficiency add up to a $7\%$ uncertainty on the
      cross sections.
\end{itemize}

The uncertainty arising from the model dependence of the acceptance
was evaluated as follows (the effect on the cross sections is shown in
parenthesis):
\begin{itemize} 
  \item the Peterson fragmentation parameter $\epsilon$ in the MC was 
        varied from $0.002$ to $0.006$ as allowed by 
        LEP and SLD measurements \cite{opal,aleph,sld}. The Lund-Bowler
        fragmentation function was used as an alternative, both
        with the default {\sc Pythia} parameters and with parameters taken 
        from OPAL measurements \cite{opal} ($\pm 2\%$);
  \item instead of using {\sc Pythia} reweighted to the NLO $p_{T}^b$
        distribution it was reweighted as a function of
        $\eta^{\rm \: jet}$ and $p^{\rm \: jet}_{T}$ 
        to agree with the measured differential distributions (-2\%) or
        without reweighting (+2\%);
  \item the fraction of flavour-excitation events in {\sc Pythia} was varied
        up and down  by a factor two ($\pm$ 4\%), as allowed by comparisons
        to the $x_{\gamma}^{\rm jets}$ distribution of the data;
  \item {\sc Herwig} was used instead of  {\sc Pythia} (-2\%).
\end{itemize}  
The total systematic uncertainty was obtained by adding the above 
contributions in quadrature. 
A 2\% overall normalization uncertainty associated with the
luminosity measurement was not included.

\section{Results}
\label{results}
All the cross sections reported below,
except for the $b$-quark cross section,
are given for dijet events with
$p_{T}^{\rm \: jet_1},p_{T}^{\rm \:jet_2}>7,6 \gev$, $\eta^{\rm \: jet_1},\eta^{\rm \: jet_2}<2.5$,
$Q^2<1 \gev^2$ and $0.2<y<0.8$.

The first set of measurements are beauty cross sections for muons
passing the cuts defined in Eq.~(\ref{e:mucuts}).
The results for the forward, barrel and rear muon-chamber
regions are shown in Fig.~\ref{f:xs1}a and Table \ref{t:xs1a}
and compared with the NLO prediction and MC models.
Both the NLO and the MC models are in reasonable agreement
 with the data. 

Figure \ref{f:xs1}b and Table \ref{t:xs1b} show the differential
 cross-section
 $d\sigma/dx_{\gamma}^{\rm jets}$
for muons in the range defined by Eq.~(\ref{e:mucuts}) in dijet events.
The $x_{\gamma}^{\rm jets}$ variable
corresponds at leading order to the fraction of the exchanged-photon
momentum in the hard scattering process.
It provides a tool to measure the relative importance of 
photon-gluon fusion, $\gamma g \rightarrow b\bbar$, which gives a peak at
  $x_{\gamma}^{\rm jets}\sim 1$,
and of other contributions, such as gluon-gluon fusion (with a gluon coming
 from the photon) or higher-order diagrams,
 which are distributed over the whole $x_{\gamma}^{\rm jets}$ range.
The sample is dominated by the high-$x_{\gamma}^{\rm jets}$ peak
 but a 
low-$x_{\gamma}^{\rm jets}$ component is also apparent. The NLO QCD
prediction describes
 the distribution well. {\sc Pythia} is also able to give a good 
description of the data due to the large contribution from
flavour excitation at low $x_{\gamma}^{\rm jets}$. {\sc Cascade},
which generates low-$x_{\gamma}^{\rm jets}$ events via initial-state
radiation without using a parton density in the
photon, tends to underestimate the cross section at
low $x_{\gamma}^{\rm jets}$.

The differential  cross sections in 
the muonic variables were measured for 
$p_{T}^{\mu}>2.5 \gev$ and $-1.6<\eta^{\mu}<2.3$. 
Figure \ref{f:xs2} and Table \ref{t:xs2}  show
 the differential cross-sections $d\sigma/d\eta^{\mu}$ 
and $d\sigma/dp_T^{\mu}$ for muons in dijet events.
The NLO QCD predictions and the MC models  
describe the $\eta^{\mu}$ distribution well. 
The $p_{T}^{\mu}$ distribution is well
reproduced by NLO QCD while the $p_{T}^{\mu}$ slope tends to be
too soft in {\sc Cascade} and {\sc Pythia}. 

The jet associated with the muon (or $\mu$-jet)
reproduces the kinematics 
of the $b$ (or $\bbar$) quark to a good approximation.
The $\mu$-jet  is defined as the jet containing
the B hadron that decays into the muon.
Figure \ref{f:xs4a}a-b and Table \ref{t:xs4aa} show the
differential cross section for the jet associated
with a muon passing the cuts of Eq.~(\ref{e:mucuts})
as a function of the jet pseudorapidity,
$d\sigma/d\eta^{\: \mu{\rm-jet}}$,
and  transverse momentum, 
$d\sigma/dp_T^{\: \mu{\rm\mbox{-}jet}}$,
for $\eta^{\mu{\rm\mbox{-}jet}}<2.5$ and $p_T^{\mu{\rm\mbox{-}jet}}>6 \gev$.
The $\mu$-jet distributions are well reproduced by the NLO
and by the MC models.

The $\mu$-jet cross sections 
have been corrected to obtain the cross sections for $b$-jets 
 in dijet events: 
$\sigma(ep \rightarrow e' \:  j j \: X)$. A $b$-jet is defined
 as a jet containing a B (or an anti-B) hadron.
This correction was performed using
{\sc Pythia} and accounts for the  $b \rightarrow \mu$ branching ratio, including direct and indirect decays, and for the full muon phase space.
Figure \ref{f:xs4a}c-d and Table \ref{t:xs4ab}
show the differential cross-sections
$d\sigma/d\eta^{\: b{\rm\mbox{-}jet}}$ and
$d\sigma/dp_{T}^{\: b{\rm\mbox{-}jet}}$. 
 The level of agreement of $b$ jets with the NLO QCD and MC predictions
 is similar to that found for the $\mu$ jets.
 It should be noted that the hadronization corrections
 in the first two $\eta^{\: b{\rm\mbox{-}jet}}$ bins are large
 ($\sim -20\%$).

 To compare the present result
 with a previous ZEUS measurement given at the $b$-quark level, the NLO
 QCD prediction was used to extrapolate the cross section for dijet
 events with a muon to the inclusive $b$-quark cross section.
 The $b$-quark differential cross section
 as a function of
 the quark transverse momentum, $d\sigma(ep \rightarrow b X)/dp_{T}^b$,
 for $b$-quark pseudorapidity in the laboratory
 frame $|\eta^b|<2$ (corresponding to a rapidity in the $ep$ centre of mass
 of $-3.75<Y^b_{\rm cms}<0.25$),
 for $Q^2<1 \gev^2$ and $0.2<y<0.8$,
 was obtained from the dijet
 cross section for events with a $\mu$-jet
 within  $|\eta^{\: \mu{\rm\mbox{-}jet}}|<2$ using the
 NLO prediction corrected for hadronization. The $\bbar$ quark
 was not considered in the definition of the $b$-quark cross section.
 As a cross-check, the measurement was corrected to the
 $b$-quark level using the {\sc Pythia} MC, giving a result in agreement
 within 6\%.
 The result, shown in Figure \ref{f:extrap} and Table \ref{t:extrap},
 is compared
 to the previous ZEUS measurement \cite{ZEUSelec}
 of the $b$-quark cross section for $p_{T}^b> p_{T}^{\rm min} = 5 \gev$
 and $|\eta^b|<2$, translated into a differential cross section
 using the NLO prediction
 and plotted at the average $b$-quark transverse momentum,
  $\langle p_{T}^b\rangle$,
 of the accepted events taken from the Monte Carlo:
 \begin{equation*}
 \bigg(\frac{d\sigma}{dp_{T}^{b}}\bigg)_{\langle p_{T}^b\rangle} =
 \bigg(\frac{d\sigma^{\rm NLO}}{dp_{T}^b}\bigg)_{\langle p_{T}^b\rangle}
 \frac{\sigma(p_{T}^b > p_{T}^{\rm min})}
      {\sigma^{\rm NLO}(p_{T}^b > p_{T}^{\rm min})}.
 \end{equation*}
 The two independent measurements are consistent and in agreement
 with the NLO QCD predictions.

\section{Conclusions}
Beauty production identified by semi-leptonic decay into muons
has been measured in dijet events
with $Q^2<1 \gev^2$. Differential cross sections for the reaction
$ep \rightarrow e^\prime \; b \bbar \; X \rightarrow e' \; j j \mu \; X'$
have been measured as a function
of the pseudorapidity and transverse momentum of the muon and of $x_{\gamma}^{\rm jets}$.
Differential cross sections for the production of
$b$-jets were also measured.

The results were compared to MC models and to a NLO QCD
prediction combined with fragmentation and B-hadron decay models.
This prediction is in good agreement with the data 
in all the differential distributions.
The {\sc Pythia} MC model is also able
to give a reasonable description of the differential cross sections.
The {\sc Cascade} MC model also gives a reasonable description of the data,
except for the  low-$x_\gamma^{\rm jets}$ region.

 The large excess of the first measurement of beauty photoproduction
 over NLO QCD, reported by the H1 
collaboration~\cite{h1bmuon1}, is not confirmed.
The present result is consistent with the previous ZEUS measurement using
semi-leptonic B decays into electrons~\cite{ZEUSelec}.
Beauty photoproduction in $ep$ collisions is reasonably well described  
both by NLO QCD and by a
MC model that includes a substantial flavour excitation
component.

\section*{Acknowledgements}
We thank the DESY Directorate for their strong support and encouragement. 
The remarkable achievements of the HERA machine group were essential for the
 successful completion of this work and are greatly appreciated.
 We are grateful for the support of the DESY computing and network services.
The design, construction and installation of the ZEUS detector have been
made possible owing to the ingenuity and effort of many people who are not
listed as authors. It is also a pleasure to thank S. Frixione and H. Jung
for help with theoretical predictions and useful conversation.

\begin{table}
\begin{center}
\begin{tabular}{|l||r|c|c||c|c|}
\hline
$\mu$-chambers & Muons & $a_{b\bbar}$ & $\sigma\pm{\rm stat.}\pm{\rm syst.}$ & $ \sigma^{NLO} \times C_{had}$ & $C_{had}$ \\ 
 & & & (pb) & (pb) & \\ \hline
 rear    &  484 & 0.15  &   $6.5\pm1.5^{+1.0}_{-1.1}$ & $4.3^{+1.6}_{-1.0}$ & $0.80$ \\
 barrel  & 2316 & 0.25  &  $38.2\pm3.4^{+5.7}_{-5.8}$ & $33.9^{+11.0}_{-7.0}$& $0.89$ \\
 forward &  868 & 0.21  &  $16.6\pm3.3^{+2.9}_{-4.6}$ & $6.5^{+2.8}_{-1.6}$& $0.86$ \\ 
\hline
\end{tabular}
\caption{\label{t:xs1a}
For each muon-chamber region defined in  Eq.~(\ref{e:mucuts})
the columns show:
the number of selected muons; the beauty
fraction $a_{b\bbar}$ obtained from the $\ptr$ fit;
the measured  beauty  cross section
with the statistical and systematic uncertainties; the NLO QCD prediction
corrected to the hadron level with the theoretical uncertainty
and the hadronization correction.
For further details see the caption to Fig.~\ref{f:xs1}.}
\end{center}
\end{table}

\begin{table}
\begin{center}
\begin{tabular}{|c|c|c|}\hline
$x_{\gamma}^{\rm jets}$ range  & $d\sigma/dx_{\gamma}^{\rm jets}\pm{\rm stat.}\pm{\rm syst.}$& $C_{had}$ \\
& (pb) & \\ \hline
0.00, 0.25  & $12.9 \pm 6.7^{+4.2}_{-7.2}$ & $0.68$ \\ 
0.25, 0.50  & $20.8 \pm 7.2^{+8.2}_{-7.7}$ & $0.83$ \\ 
0.50, 0.75  & $30.0 \pm 5.2^{+6.4}_{-4.1}$ & $0.86$ \\ 
0.75, 1.00  & $165 \pm 14^{+22}_{-34}$     & $0.90$ \\ \hline 
\end{tabular}
\caption{\label{t:xs1b}
Differential muon cross section as a function of
$x_{\gamma}^{\rm jets}$.
The multiplicative hadronization correction
applied to the NLO prediciton is shown in the last column.
For further details see the caption to Fig.~\ref{f:xs1}.}
\end{center}
\end{table}

\begin{table}
\begin{center}
\begin{tabular}{|c|c|c|}\hline
$\eta^{\mu}$ range  & $d\sigma/d\eta^{\mu}\pm{\rm stat.}\pm{\rm syst.}$  & $C_{had}$ \\
  & (pb) & \\ \hline
-1.6, -0.75 & $4.6 \pm 1.3^{+0.8}_{-0.9}$ & $0.83$ \\ 
-0.75, 0.25 &$18.8 \pm 2.2^{+2.7}_{-2.7}$ & $0.88$ \\ 
 0.25, 1.30 &$16.7 \pm 2.4^{+2.7}_{-2.9}$ & $0.92$ \\ 
 1.30, 2.30 &$10.0 \pm 2.3^{+1.4}_{-2.6}$ & $0.91$ \\ \hline \hline
$p_{T}^{\mu}$ range & $d\sigma/dp_{T}^{\mu}\pm{\rm stat.}\pm{\rm syst.}$   & $C_{had}$ \\
 (GeV)  & (pb/GeV) & \\ \hline
2.5, 4.0  &$16.1  \pm 2.3^{+2.9}_{-3.7}$ & $0.87$ \\ 
4.0, 6.0  &$ 7.1  \pm 1.0^{+1.3}_{-1.2}$ & $0.92$ \\
6.0, 10.0 &$ 1.69 \pm 0.32^{+0.22}_{-0.24}$ & $0.97$ \\ \hline
\end{tabular}
\caption{\label{t:xs2}
Differential muon cross section as a function of
$\eta^{\mu}$ and $p_{T}^{\mu}$.
For further details see the caption to Fig.~\ref{f:xs2}.}

\end{center}
\end{table}

\begin{table}
\begin{center}
\begin{tabular}{|c|c|c|} \hline
$\eta^{\mu{\rm \mbox{-}jet}}$ range & $d\sigma/d\eta^{\mu{\rm \mbox{-}jet}}\pm{\rm stat.}\pm{\rm syst.}$  & $C_{had}$ \\ 
 & (pb) &  \\ \hline
-1.6,-0.6  &  $8.6 \pm 1.7^{+1.4}_{-1.8}$ & $0.73$ \\
-0.6, 0.4  &  $18.9 \pm 2.2^{+2.8}_{-2.7}$ & $0.84$  \\
0.4, 1.4  & $14.3 \pm 2.3^{+2.5}_{-3.8}$ & $0.97$ \\
1.4, 2.5   &  $14.4 \pm 2.7^{+2.3}_{-4.2}$ & $1.00$ \\ \hline \hline
$p_{T}^{\mu{\rm \mbox{-}jet}}$ range & $d\sigma/dp_{T}^{\mu{\rm \mbox{-}jet}}\pm{\rm stat.}\pm{\rm syst.}$  & $C_{had}$ \\
(GeV) & (pb/GeV) & \\ \hline
6,11 & $6.41 \pm 0.67^{+1.00}_{-1.24}$ & $0.88$ \\
11,16 & $2.98 \pm 0.37^{+0.52}_{-0.82}$ & $0.88$ \\
16,30 & $0.51 \pm 0.11^{+0.08}_{-0.08}$   & $0.90$ \\ \hline
\end{tabular}
\caption{\label{t:xs4aa}
Differential cross section for jets associated with
a muon as a function of $\eta^{\mu{\rm \mbox{-}jet}}$
and $p_{T}^{\mu{\rm \mbox{-}jet}}$.
For further details see the caption to Fig.~\ref{f:xs4a}.}
\end{center}
\end{table}

\begin{table}
\begin{center}
\begin{tabular}{|c|c|c|} \hline
$\eta^{b{\rm \mbox{-}jet}}$ range & $d\sigma/d\eta^{b{\rm\mbox{-}jet}}\pm{\rm stat.}\pm{\rm syst.}$ & $C_{had}$ \\
 & (pb) & \\ \hline
-1.6,-0.6 & $ 152 \pm 29^{+24}_{-31}$  & $0.68$  \\ 
-0.6, 0.4  &$ 356 \pm 41^{+59}_{-53}$ & $0.78$ \\ 
0.4, 1.4 &  $ 275 \pm 45^{+53}_{-73}$     & $0.96$ \\ 
1.4, 2.5 & $ 229 \pm 44^{+41}_{-69}$     & $1.06$ \\ \hline \hline
$p_{T}^{b{\rm\mbox{-}jet}}$ range & $d\sigma/dp_{T}^{b{\rm\mbox{-}jet}}\pm{\rm stat.}\pm{\rm syst.}$  & $C_{had}$ \\
(GeV) & (pb/GeV) & \\ \hline
6,11 & $137 \pm 14^{+21}_{-27}$  & $0.85$ \\ 
11,16& $ 43.8 \pm 5.5^{+7.7}_{-12.0}$ & $0.86$ \\  
16,30 &  $ 5.7 \pm 1.2^{+1.0}_{-0.9}$   & $0.89$ \\ \hline
\end{tabular}
\caption{\label{t:xs4ab}
Differential cross section for $b$-jets
as a function of $\eta^{b{\rm \mbox{-}jet}}$
and $p_{T}^{b{\rm \mbox{-}jet}}$.
For further details see the caption to Fig.~\ref{f:xs4a}.}
\end{center}
\end{table}

\begin{table}
\begin{center}
\begin{tabular}{|c|c|c|c|c|} \hline
$p_{T}^{\mu{\rm\mbox{-}jet}}$ range & $d\sigma/dp_{T}^{\mu{\rm \mbox{-}jet}}\pm{\rm stat.}\pm{\rm syst.}$  & $C_{had}$ & $\langle p_{T}^b \rangle$ & $d\sigma/dp_{T}^b \pm{\rm stat.}\pm{\rm syst.}$  \\ 
(GeV) & ($|\eta^{\mu{\rm\mbox{-}jet}}|<2$) (pb/GeV) & & (GeV) & (pb/GeV) \\ \hline
6,11  & $6.22 \pm 0.63^{+0.91}_{-1.04}$ & $0.87$ &  8.5 & $90 \pm 9 ^{+13}_{-15}$\\ 
11,16 & $2.74 \pm 0.36^{+0.47}_{-0.74}$ & $0.88$ & 13.6 & $18.2 \pm 2.4 ^{+3.1}_{-4.9}$\\ 
16,30 & $0.43 \pm 0.10^{+0.07}_{-0.06}$ & $0.88$ & 21.25 & $2.0 \pm 0.5 ^{+0.3}_{-0.3}$\\ \hline
\end{tabular}
\caption{\label{t:extrap}
Differential $b$-quark cross-section $d\sigma/dp_{T}^b$.
The first columns show the differential $\mu$-jet cross-section 
$d\sigma/dp_{T}^{\mu{\rm \mbox{-}jet}}$
restricted to $|\eta^{\mu{\rm\mbox{-}jet}}|<2$, 
the corresponding hadronization correction and
the average $b$-quark transverse momentum,
 $\langle p_{T}^b \rangle$, as obtained from the {\sc Pythia} MC. The 
 $b$-quark differential cross section $d\sigma/dp_T^b$   for $|\eta^b|<2$,
 evaluated at $\langle p_{T}^b \rangle$,
 is shown in the last column.
For further details see the caption to Fig.~\ref{f:extrap}.}
\end{center}
\end{table}

{
\def\bibname{\Large\bf References}
\def\refname{\Large\bf References}
\pagestyle{plain}
\ifzeusbst
  \bibliographystyle{./BiBTeX/bst/l4z_default}
\fi
\ifzdrftbst
  \bibliographystyle{./BiBTeX/bst/l4z_draft}
\fi
\ifzbstepj
  \bibliographystyle{./BiBTeX/bst/l4z_epj}
\fi
\ifzbstnp
  \bibliographystyle{./BiBTeX/bst/l4z_np}
\fi
\ifzbstpl
  \bibliographystyle{./BiBTeX/bst/l4z_pl}
\fi
{\raggedright
\bibliography{./BiBTeX/user/syn,%
              ./BiBTeX/bib/l4z_articles.bib,%
              ./BiBTeX/bib/l4z_books.bib,%
              ./BiBTeX/bib/l4z_conferences.bib,%
              ./BiBTeX/bib/l4z_h1.bib,%
              ./BiBTeX/bib/l4z_misc.bib,%
              ./BiBTeX/bib/l4z_old.bib,%
              ./BiBTeX/bib/l4z_preprints.bib,%
              ./BiBTeX/bib/l4z_replaced.bib,%
              ./BiBTeX/bib/l4z_temporary.bib,%
              ./BiBTeX/bib/l4z_zeus.bib}}
}
\vfill\eject

%

\begin{figure}[t]
\centerline{{  \leavevmode     \epsfxsize=8cm \epsffile{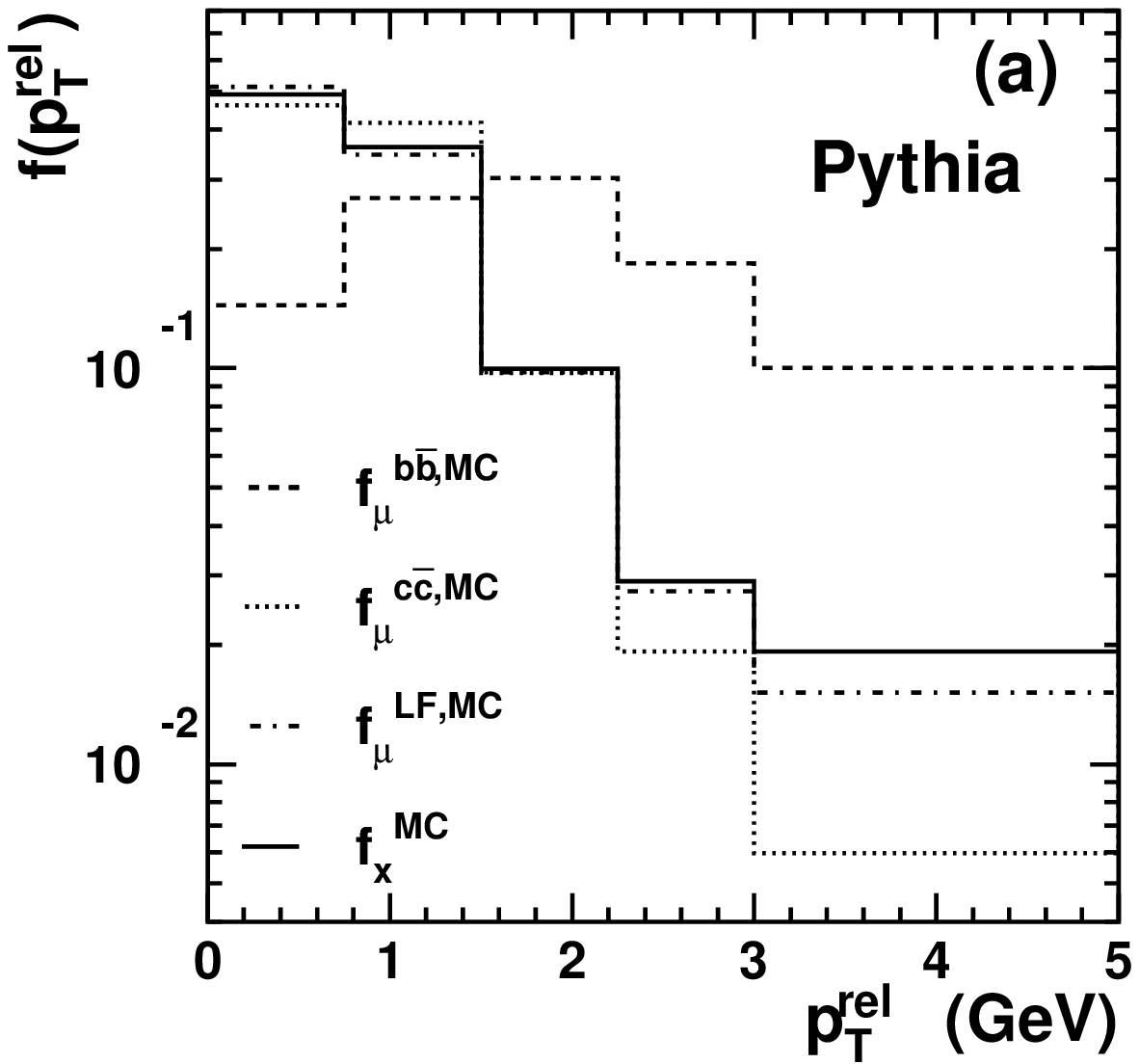} 
            \hspace*{-0.8 cm}  \epsfxsize=8cm \epsffile{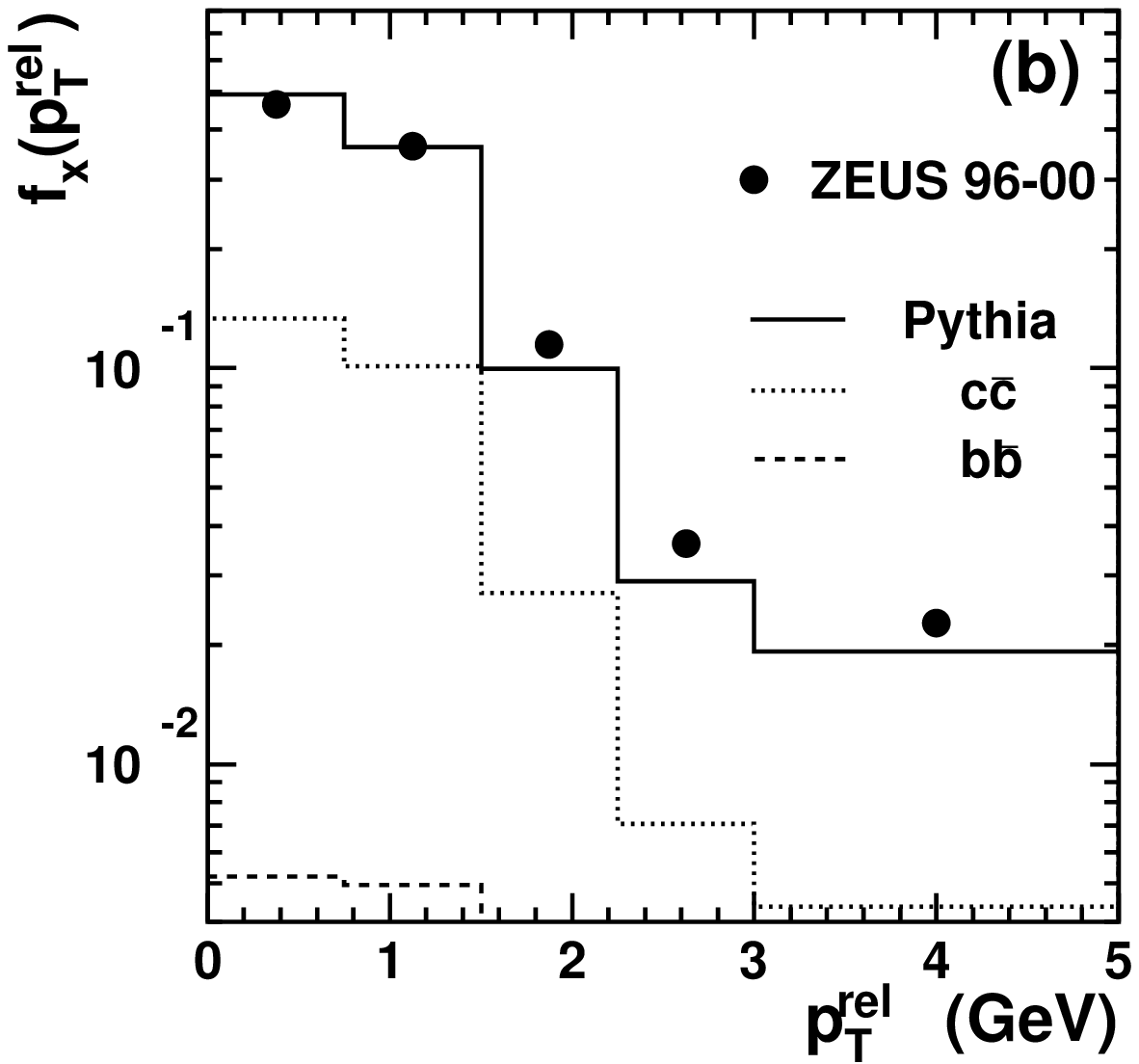} }}
\caption[this space for rent]{
(a) The $\ptr$ distribution as predicted by
 the {\sc Pythia} Monte Carlo for
  reconstructed muons from beauty ($f_{\mu}^{b\bbar,{\rm MC}}$, dashed histogram), charm  ($f_{\mu}^{c\cbar,{\rm MC}}$, dotted histogram)
and light-flavours ($f_{\mu}^{\rm LF,MC}$, dash-dotted histogram),
 and for  unidentified  tracks ($f_{x}^{\rm MC}$, full-line histogram).
 The distributions are normalized to unity.
 (b) The $\ptr$ distribution of unidentified data tracks
  (points), compared to the prediction from
 {\sc Pythia} (full line).
 The  charm  and beauty components 
are also shown as the dotted- and dashed-line  histograms, respectively.}
\label{f:ptr0}
\end{figure}

\begin{figure}[t]
\centerline{ \epsfxsize=14cm \epsffile{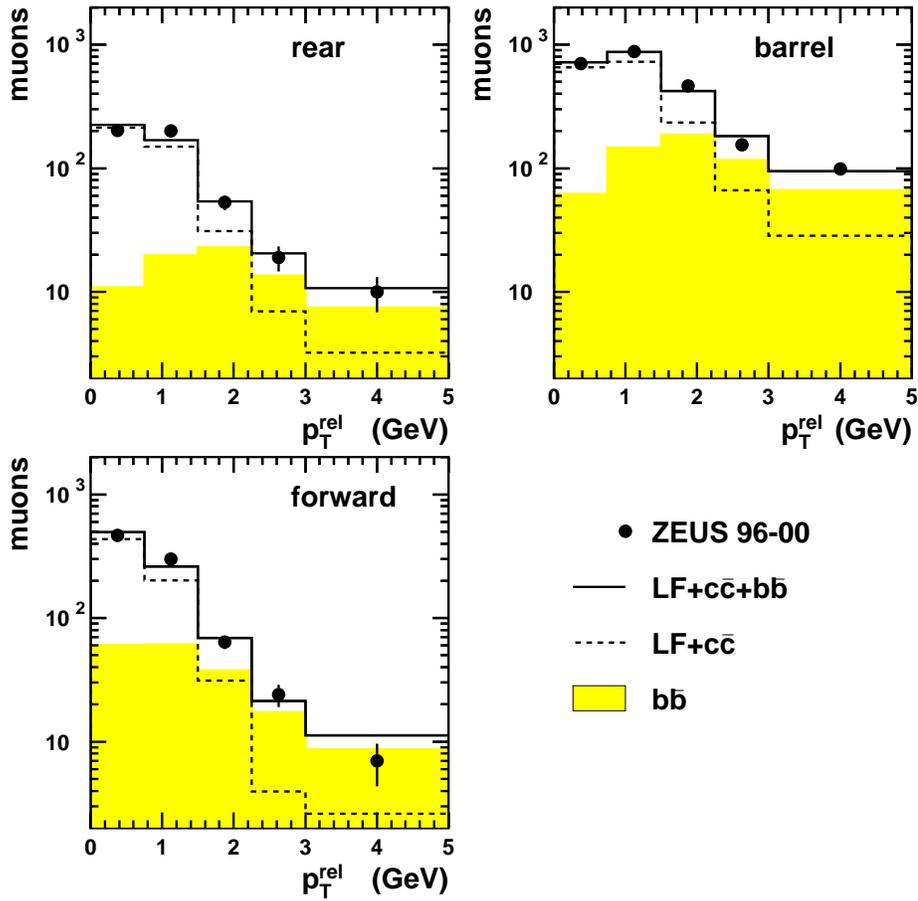}}
\caption[this space for rent]{The $\ptr$ distribution for events with muons in the rear, barrel and forward regions defined in Eq.~(\ref{e:mucuts}).
 The data (points) are compared to the mixture of beauty and charm+LF background obtained from the $\ptr$ fit (full line).
 The shaded histogram represents the beauty component while the dashed-line
 histogram is the background.}
\label{f:fit}
\end{figure}

\begin{figure}[t]
\centerline{{ \leavevmode \epsfxsize=16cm \epsffile{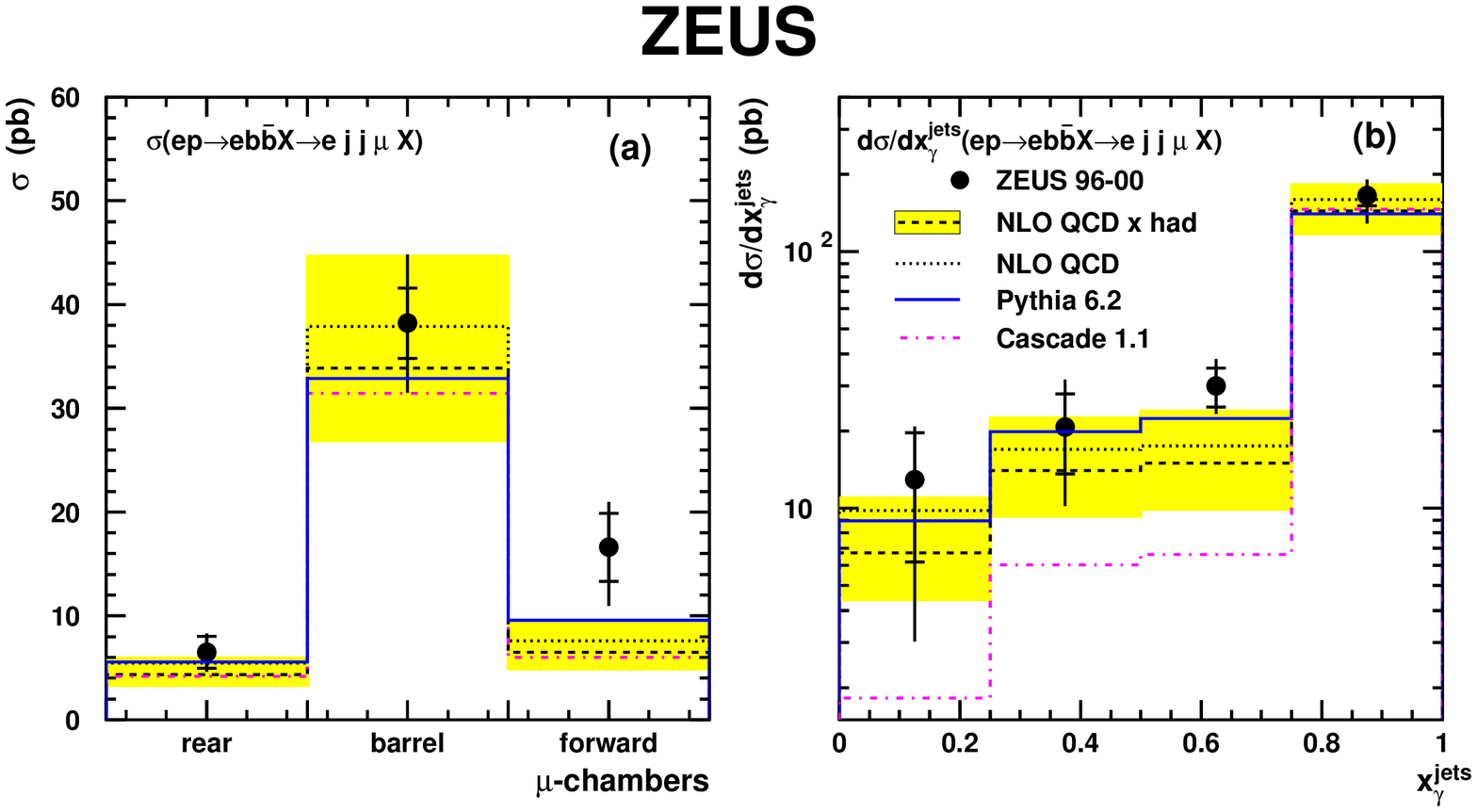}}}
\caption[this space for rent]{
Cross sections for muons coming from
$b$ decays in dijet events with 
$p_{T}^{\rm jet_1}>7,p_{T}^{\rm jet_2}>6 \gev$,
$\eta^{\rm jet_1},\eta^{\rm jet_2} <2.5$, 
$0.2<y<0.8$, $Q^2<1\gev^2$
passing the selection of Eq.~(\ref{e:mucuts}).
(a) The cross section for the forward, barrel and rear regions
(defined in  Eq.~(\ref{e:mucuts})). (b)
The differential cross section as a function of $x_\gamma^{\rm jets}$.
 The data (points) are compared to 
the predictions of NLO QCD (dotted line: parton-level jets; 
dashed line: jets corrected to the hadron level).
The full error bars are the quadratic sum of the statistical (inner part) and systematic uncertainties. The band around the NLO prediction represents the variation on the theoretical predictions obtained by varying the $b$-quark mass, $\mu_r$ and $\mu_f$, as explained in the text. The data are also compared to the predictions of the {\sc Pythia} (solid line histogram) and {\sc Cascade} (dot-dashed line histogram) Monte Carlo models.}
\label{f:xs1}
\end{figure}

\begin{figure}[t]
\centerline{{ \leavevmode \epsfxsize=16cm \epsffile{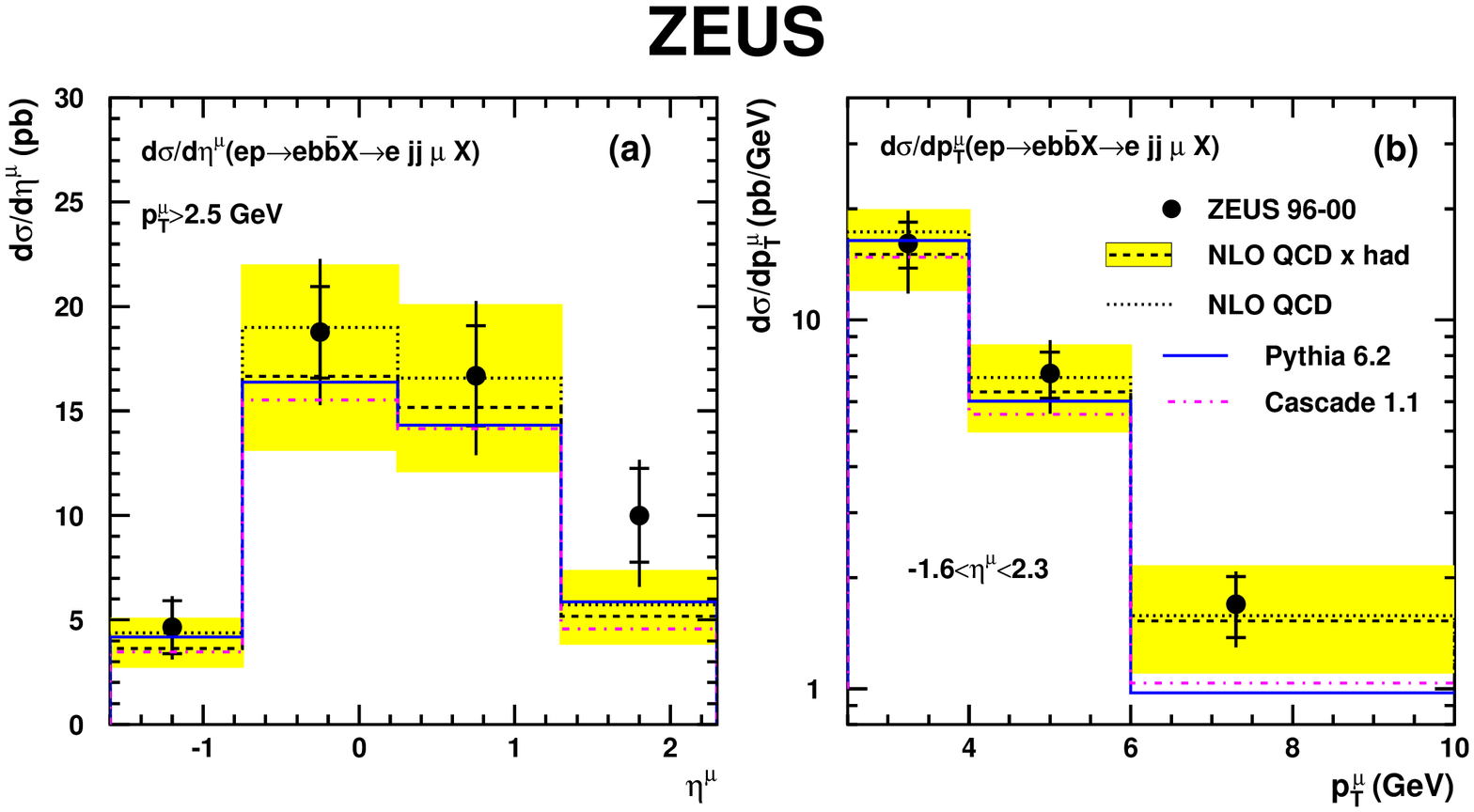}}}
\caption[this space for rent]{
Differential cross sections 
as a function of (a) the muon pseudorapidity $\eta^\mu$ and
(b) transverse momentum $p_{T}^\mu$, for $p_{T}^\mu>2.5\gev$
and $-1.6<\eta^{\mu}<2.3$, 
for muons coming from
$b$ decays in dijet events with 
$p_{T}^{\rm jet_1}>7, p_{T}^{\rm jet_2}>6 \gev$,
$\eta^{\rm jet_1},\eta^{\rm jet_2} <2.5$, 
$0.2<y<0.8$, $Q^2<1\gev^2$.
The data (points) are compared to the predictions of NLO QCD (dotted line: parton-level jets; dashed line: corrected to hadron-level jets). The full error bars are the quadratic sum of the statistical (inner part) and systematic uncertainties. The band around the NLO prediction represents the variation on the theoretical predictions obtained by varying the $b$-quark mass and $\mu_r$ and $\mu_f$ as explained in the text.The data are also compared to the predictions of the {\sc Pythia} (solid line histogram) and {\sc Cascade} (dot-dashed line histogram) Monte Carlo models.}
\label{f:xs2}
\end{figure}

\begin{figure}[t]
\centerline{{ \leavevmode \epsfxsize=16cm \epsffile{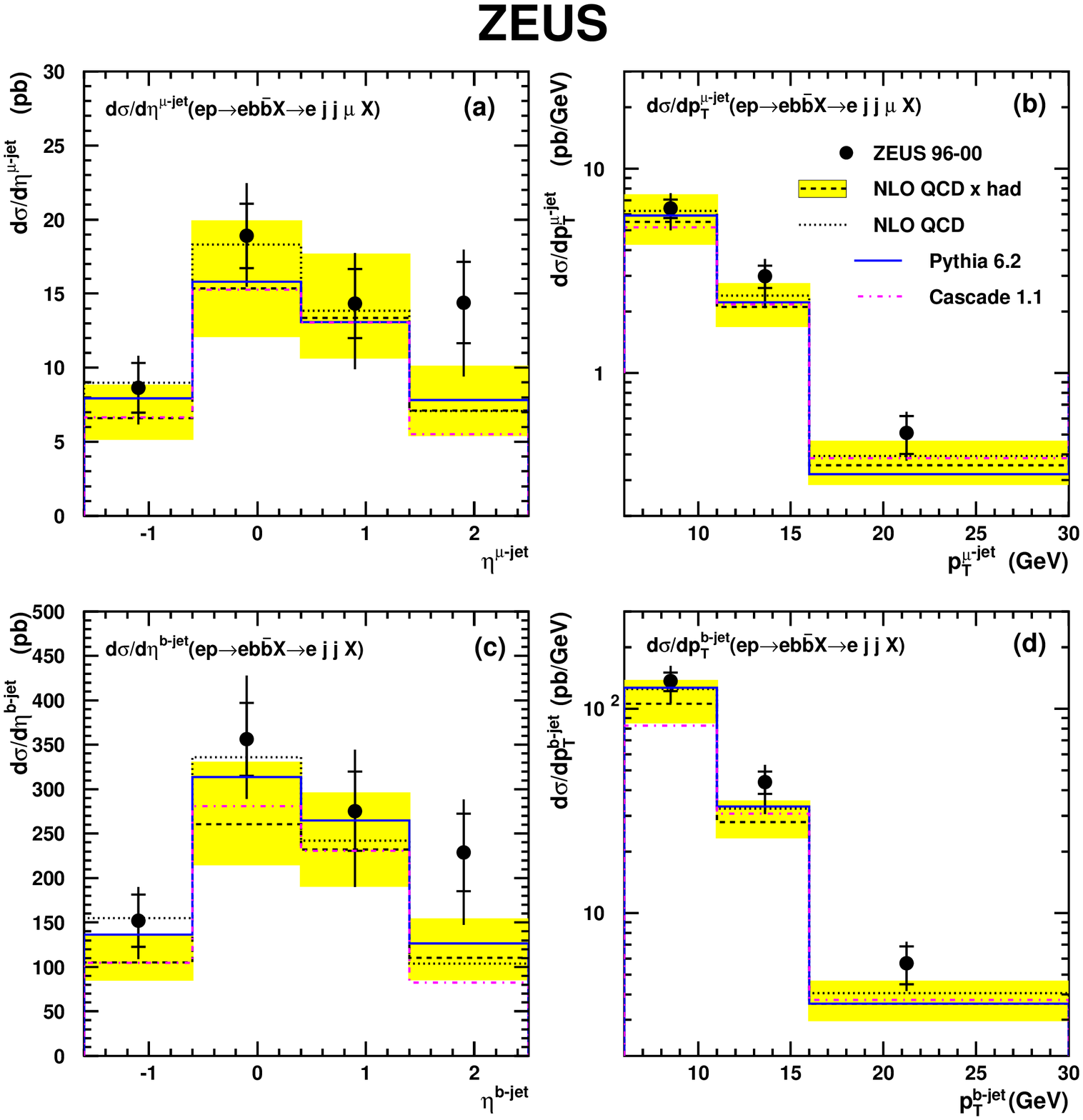}}}
\caption[this space for rent]{
Differential cross sections 
as a function of (a) the pseudorapidity $\eta^{\rm \mu-jet}$ and (b)
the transverse momentum $p_{T}^{\rm \mu-jet}$ of the jet associated 
to the muon, for $p_{T}^{\rm \mu-jet}>6\gev$, 
$\eta^{\rm \mu-jet}<2.5$,
for muons passing the selection of Eq.~(\ref{e:mucuts}) and 
coming from $b$ decays; differential cross sections as a 
function of (c) the pseudorapidity $\eta^{b-{\rm jet}}$
 and (d) the transverse momentum $p_{T}^{b-{\rm jet}}$ of the jet
containing a B hadron. All the cross sections are evaluated for 
 dijet events with 
$p_{T}^{\rm jet_1}>7, p_{T}^{\rm jet_2}>6 \gev$,
$\eta^{\rm jet_1},\eta^{\rm jet_2} <2.5$, 
$0.2<y<0.8$, $Q^2<1\gev^2$.
 The data (points) are compared to the predictions of NLO QCD
  (dotted line: parton level;  dashed line: corrected to hadron level).
  The full error bars are the quadratic sum of the statistical (inner part)
 and systematic uncertainties. The band around the NLO prediction 
represents the  uncertainty on the theoretical prediction 
corrected for hadronization.
  The data are also compared to the predictions of the {\sc Pythia}
 (solid line histogram) and {\sc Cascade} (dot-dashed line histogram)
 Monte Carlo models.}
\label{f:xs4a}
\end{figure}


\begin{figure}[t]
\centerline{{ \leavevmode \epsfxsize=13cm \epsffile{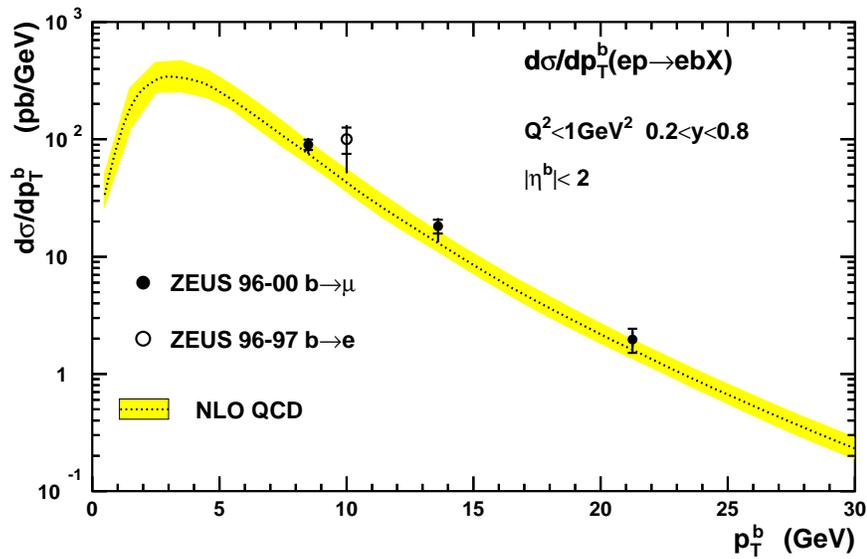}}}
\caption[this space for rent]{Differential cross section 
for $b$-quark production as a function of the $b$-quark
transverse momentum $p_{T}^b$ for $b$-quark
pseudorapidity $|\eta^b|<2$ and for $Q^2<1 \gev^2$, $0.2<y<0.8$.
The filled points show the ZEUS results from this
analysis and the open point is the previous ZEUS measurement 
in the electron channel \cite{ZEUSelec}. 
The full error bars are the quadratic sum of the statistical (inner part) and
 systematic uncertainties. The dashed line
shows the NLO QCD prediction with the theoretical uncertainty shown as the
 shaded band.}
\label{f:extrap}
\end{figure}

%
\end{document}